\documentclass[a4paper,11pt]{article}
\usepackage{jinstpub}

\usepackage[T5,T1]{fontenc}
\usepackage[mathlines]{lineno}
\usepackage{etoolbox}
\usepackage{cleveref}
\makeatletter
\patchcmd{\@floatboxreset}{\reset@font}{\reset@font\linenumbers}{}{}
\makeatother
\usepackage{booktabs}
\usepackage{float}
\usepackage{graphicx}
\usepackage{subcaption}

\makeatletter
\newcommand{\supplementpagestyle}{%
  \def\ps@suppl{%
    \def\@oddhead{}%
    \def\@evenhead{}%
    \def\@oddfoot{\hfil\thepage\hfil}%
    \def\@evenfoot{\hfil\thepage\hfil}%
  }%
}
\makeatother

\newcommand{\DynEdge}{\texttt{DynEdge}}
\newcommand{\WavePID}{\texttt{WavePID}}
\newcommand{\Geant}{\texttt{Geant4}}

\title{\boldmath WavePID: Low-energy flavor identification using single-PMT time series in IceCube}

\author[15]{R. Abbasi,}
\author[62]{M. Ackermann,}
\author[16]{J. Adams,}
\author[9]{J. A. Aguilar,}
\author[20]{M. Ahlers,}
\author[21]{J.M. Alameddine,}
\author[34]{S. Ali,}
\author[42]{N. M. Amin,}
\author[40]{K. Andeen,}
\author[12]{C. Arg{\"u}elles,}
% ...
\affiliation[9]{Universit{\'e} Libre de Bruxelles, Science Faculty CP230, B-1050 Brussels, Belgium}
% ...
\affiliation[12]{Department of Physics and Laboratory for Particle Physics and Cosmology, Harvard University, Cambridge, MA 02138, USA}
% ...
\affiliation[15]{Department of Physics, Loyola University Chicago, Chicago, IL 60660, USA}
\affiliation[16]{Dept. of Physics and Astronomy, University of Canterbury, Private Bag 4800, Christchurch, New Zealand}
% ...
\affiliation[20]{Niels Bohr Institute, University of Copenhagen, DK-2100 Copenhagen, Denmark}
\affiliation[21]{Dept. of Physics, TU Dortmund University, D-44221 Dortmund, Germany}
% ...
\affiliation[34]{Dept. of Physics and Astronomy, University of Kansas, Lawrence, KS 66045, USA}
% ...
\affiliation[40]{Department of Physics, Marquette University, Milwaukee, WI 53201, USA}
% ...
\affiliation[42]{Bartol Research Institute and Dept. of Physics and Astronomy, University of Delaware, Newark, DE 19716, USA}
% ...
\affiliation[62]{Deutsches Elektronen-Synchrotron DESY, Platanenallee 6, D-15738 Zeuthen, Germany}
% ...

\abstract{
The IceCube Neutrino Observatory, a cubic-kilometer detector at the South Pole, identifies neutrino flavor through event morphology. 
Sparse photon detection makes this classification particularly challenging in the 5--100~GeV regime, the energy range relevant for oscillation measurements and searches for physics beyond the Standard Model. 
We introduce \WavePID, a template-based log-likelihood-ratio classifier that exploits nanosecond-scale timing on individual detector modules through three observables:
the distance to the reconstructed vertex, the early-charge fraction, and the module-to-module time difference. 
Evaluated on a cascade-enriched sample selected by a state-of-the-art graph neural network, \WavePID\ improves both cascade purity and classification performance over the neural network alone. 
This demonstrates that per-module pulse timing carries flavor-identification information complementary to morphology-based classifiers, opening a new physics-motivated observable for low-energy neutrino reconstruction. 
\Geant\ simulations associate this signal with differences in Cherenkov emission geometry between muon tracks and electromagnetic showers. 
These results motivate exploiting nanosecond-scale pulse timing in future low-energy classifiers and in detector designs with improved per-module timing in next-generation neutrino telescopes.}


\usepackage{caption}

\begin{document}
\maketitle

\flushbottom
%\linenumbers
%------------------------------------------------------------------------------
\section{Introduction}
\label{sec:intro}
\subsection{The IceCube Detector}
The IceCube Neutrino Observatory~\cite{IceCube2017JINST} is a cubic-kilometer neutrino detector instrumented with optical sensors called Digital Optical Modules (DOMs) embedded in the deep, glacial ice at the geographic South Pole.
It features a densely instrumented sub-array, DeepCore, designed for low-energy event reconstruction, extending IceCube's sensitivity down to several GeV~\cite{IceCube:2011ucd}.
This extended energy range enables precision measurements of atmospheric oscillation parameters and searches for physics beyond the Standard Model~\cite{IceCube:2011ucd,DeepCoreOsc2023,DeepCoreMassOrdering2020,DeepCoreSterile2024}.
Flavor separation is central to this scientific case, and it is the primary focus of the method that we introduce in this article.

Neutrino interactions in or near the instrumented volume produce charged secondaries that emit Cherenkov photons, which propagate through the ice and are detected by the DOMs.
Depending on the final-state particle content, two distinct event morphologies are defined.
\emph{Track} events arise when a relativistic muon traverses the detector, producing an elongated light pattern.
This morphology is associated primarily with charged-current (CC) $\nu_\mu$ interactions and, less frequently, CC $\nu_\tau$ interactions in which the $\tau$ decays to a $\mu$ and neutrinos.
\emph{Cascade} events correspond to localized, approximately isotropic light depositions from CC $\nu_e$ and $\nu_\tau$ interactions and neutral-current interactions of all flavors.
Track--cascade separation enters atmospheric mixing analyses~\cite{DeepCoreOsc2023, IceCube:2024nhk}, the neutrino mass
ordering~\cite{DeepCoreMassOrdering2020}, $\nu_\tau$
appearance~\cite{IceCube:2019dqi}, and sterile neutrino
searches~\cite{DeepCoreSterile2024}. Misclassified events migrate between
flavor channels and inflate the uncertainties on the extracted
parameters.

For detection, each DOM houses a photomultiplier tube (PMT) that records time-resolved waveforms. Pulses read out by the Analog Transient Waveform Digitizer (ATWD) are sampled in 128 bins of 3.3~ns width, providing nanosecond-level timing information. Calibrated pulse series are extracted from the recorded waveforms by deconvolution of a single photoelectron template~\cite{IceCube2017JINST}.

\subsection{The challenge of low-energy flavor identification}
\begin{figure}[htbp]
  
  \centering
  \includegraphics[width=1\linewidth]{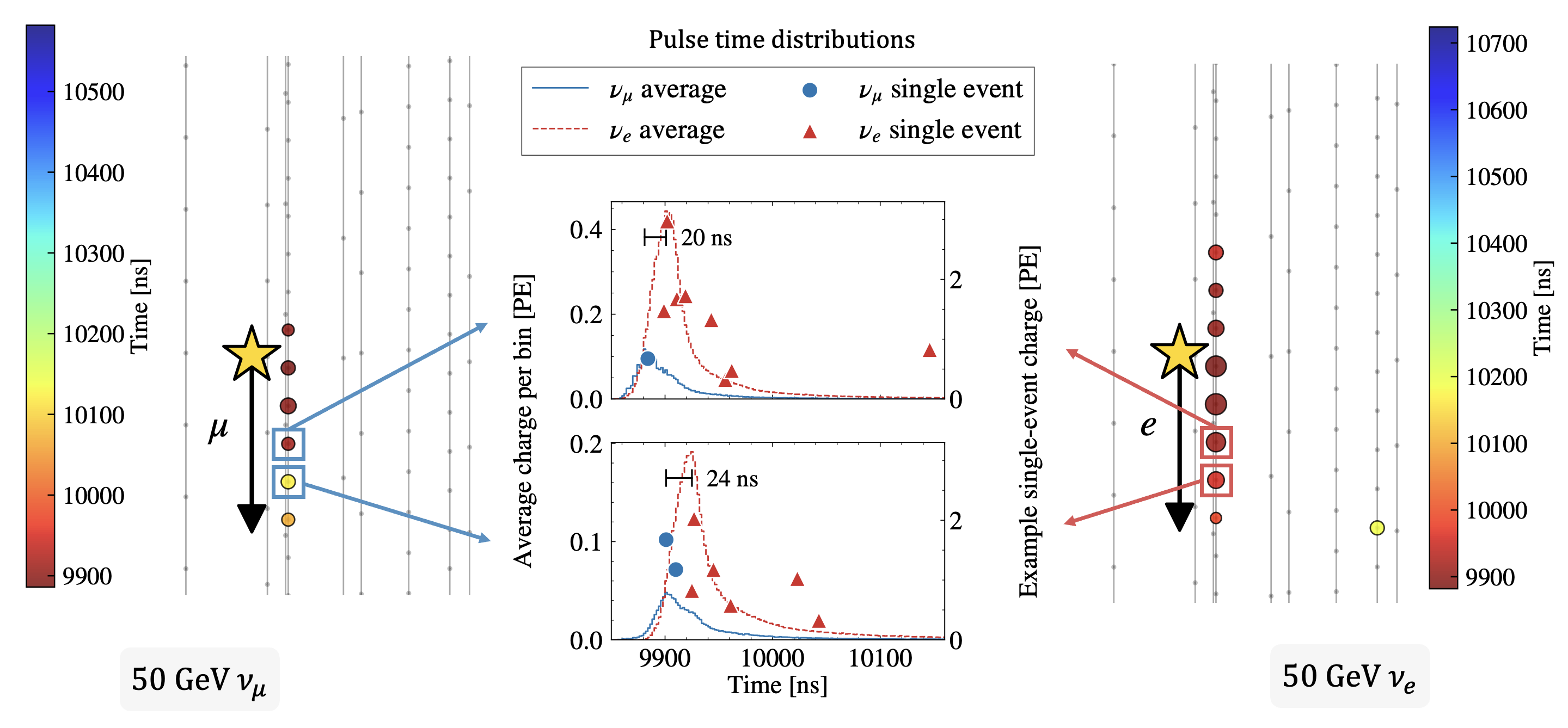}%
  
\caption{
Neutrino event signatures: 50~GeV $\nu_\mu$ CC (left) and $\nu_e$ CC (right) interactions, with 35~GeV deposited by the charged lepton and 15~GeV by hadrons. A star marks the neutrino interaction vertex, sphere size encodes charge, and color encodes timing. At this hit multiplicity (13 for $\nu_\mu$, 18 for $\nu_e$), the morphology provides limited discrimination. The center panel shows pulse time distributions from $10^5$ photon propagations for two selected DOMs, with horizontal bars marking the peak separation that \WavePID\ exploits. Example pulse series are overlaid as markers.
%The 15--20~ns peak separation, comparable to the distribution widths, provides clear per-DOM discrimination.
}
  \label{fig:event_displays}

\end{figure}

Dim events with sparse light detection are common in the 5--100~GeV range, the low-energy end of IceCube's GeV--PeV sensitivity, posing the classification challenge this study addresses.
The event-level hit multiplicity \(n_{\mathrm{hit}}\) is the number of DOMs recording at least one pulse after noise cleaning~\cite{LowEnergyReco2022}.
This study focuses on the regime $10 \leq n_\mathrm{hit} < 20$, chosen because (i) it contains over half of detected low-energy neutrino events in IceCube DeepCore MC; (ii) it aligns with the standard low-energy event selection used in oscillation analyses; and (iii) it is the regime where timing-based methods offer the largest relative gain, because the light pattern weakly constrains the event morphology, as illustrated in \cref{fig:event_displays}.

State-of-the-art low-energy classification in IceCube uses graph neural networks (GNNs) such as \DynEdge~\cite{DynEdge2022}. While this particle identification (PID) method performs well for events with many hits, the achievable separation power decreases sharply at lower hit multiplicities, motivating a more effective use of fine-grained timing information.
Next-generation neutrino telescopes increasingly build improved per-module timing into the optical-module architecture by design~\cite{KM3NeT_multiPMT_2022, TRIDENT_NatAstron_2023, POne_2020, HUNT_ICRC2023, IceCubeGen2_2021}, replacing the single large photomultiplier per module with smaller multi-PMT arrays or silicon photomultipliers and faster digitization. This is especially powerful in water, where the longer scattering length preserves the fine timing structure at the sensor.
The central question of this work is whether nanosecond-scale timing on individual DOMs carries PID information that existing classifiers do not capture.
Because a muon track produces a sharp, direct Cherenkov front while an electromagnetic cascade builds up its light from secondaries over several nanoseconds (cf. \cref{sec:Geant4}), the discriminating structure is concentrated in a brief $\mathcal{O}(10\,\mathrm{ns})$ window following the first pulse on each DOM.
Although \DynEdge\ ingests the raw per-pulse information (DOM position, time, and charge) as node features and learns its own representation from it, the pulse timestamps are normalized to the $\mathcal{O}(\mu\mathrm{s})$ event duration, which may compress this early-time structure into a small fraction of the node feature range.
\WavePID, on the other hand, operates on per-DOM pulse series data and explicitly targets the relative charge accumulation within the first several nanoseconds.

This early-time regime is motivated by the demonstrated performance of waveform-based PID in single PMTs in controlled water Cherenkov measurements.
Specifically, in the Fermilab FNAL-1267 test-beam study~\cite{Samani2020}, pulse-shape information from a single PMT was used to separate minimum ionizing pions from radiative electrons in the several-GeV regime.
The work presented here tests whether the discrimination demonstrated at Fermilab transfers to IceCube, representing—to our knowledge—the first explicit use of single-DOM nanosecond-scale timing information for PID in a neutrino telescope.

The remainder of this article is organized as follows. \Cref{sec:Geant4} 
presents \Geant\ photon arrival-time studies investigating the underlying 
microphysics and motivating the timing observables. \Cref{sec:wavepid} 
introduces \WavePID\ as a compact classifier that exploits pulse 
timing on the order of ten nanoseconds as an additional classification dimension, complementing existing 
morphology-based methods in the low-hit regime. \Cref{sec:results} evaluates 
its performance and systematics on IceCube DeepCore MC in the sparse target 
regime $n_{\mathrm{hit}} \in [10,20)$, and \cref{sec:conclusion} concludes.
%------------------------------------------------------------------------------
\section{The origin of photon hits on DOMs}
\label{sec:Geant4}
Motivated by particle-dependent waveform differences observed with a single PMT in water~\cite{Samani2020}, we first investigate whether analogous timing differences arise between muons and electromagnetic showers in ice and which processes generate them. To this end, we simulate photon arrival time distributions (PATDs) on a single DOM using \Geant~\cite{Geant42003}, comparing muons and electrons at common primary energies and impact parameters. Holding these quantities fixed provides a controlled comparison in which differences in the PATDs can be associated with particle type and the corresponding light-production mechanisms. This controlled single-DOM study provides the microphysics motivation for \WavePID, while the detector-level question of where this timing information is useful in addition to event morphology is addressed with the full IceCube simulation in \cref{sec:results}.

\subsection{\Geant\ simulation setup}
\begin{figure}[htbp]
  \centering
  \includegraphics[width=0.47\linewidth]{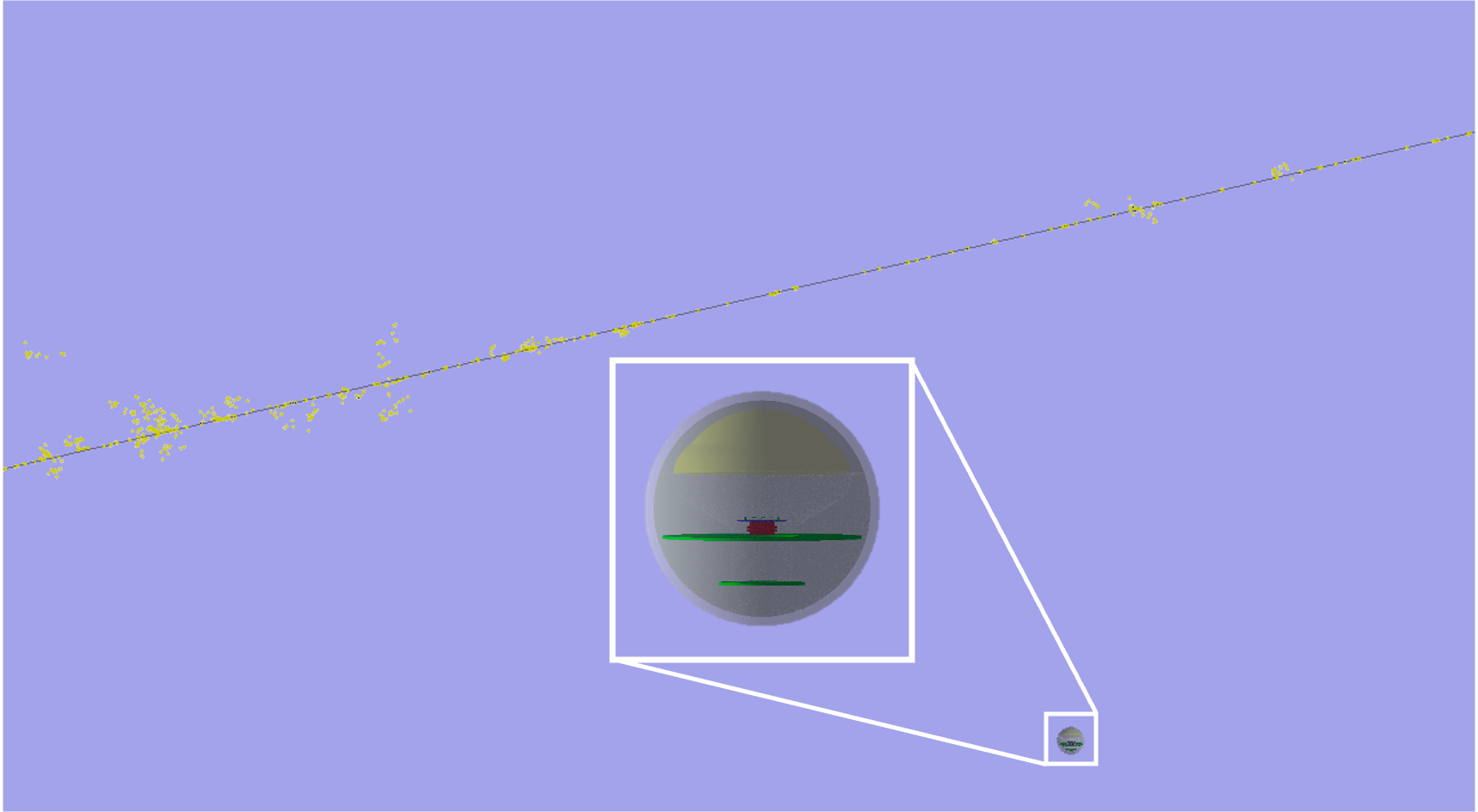}\hfill
  \includegraphics[width=0.47\linewidth]{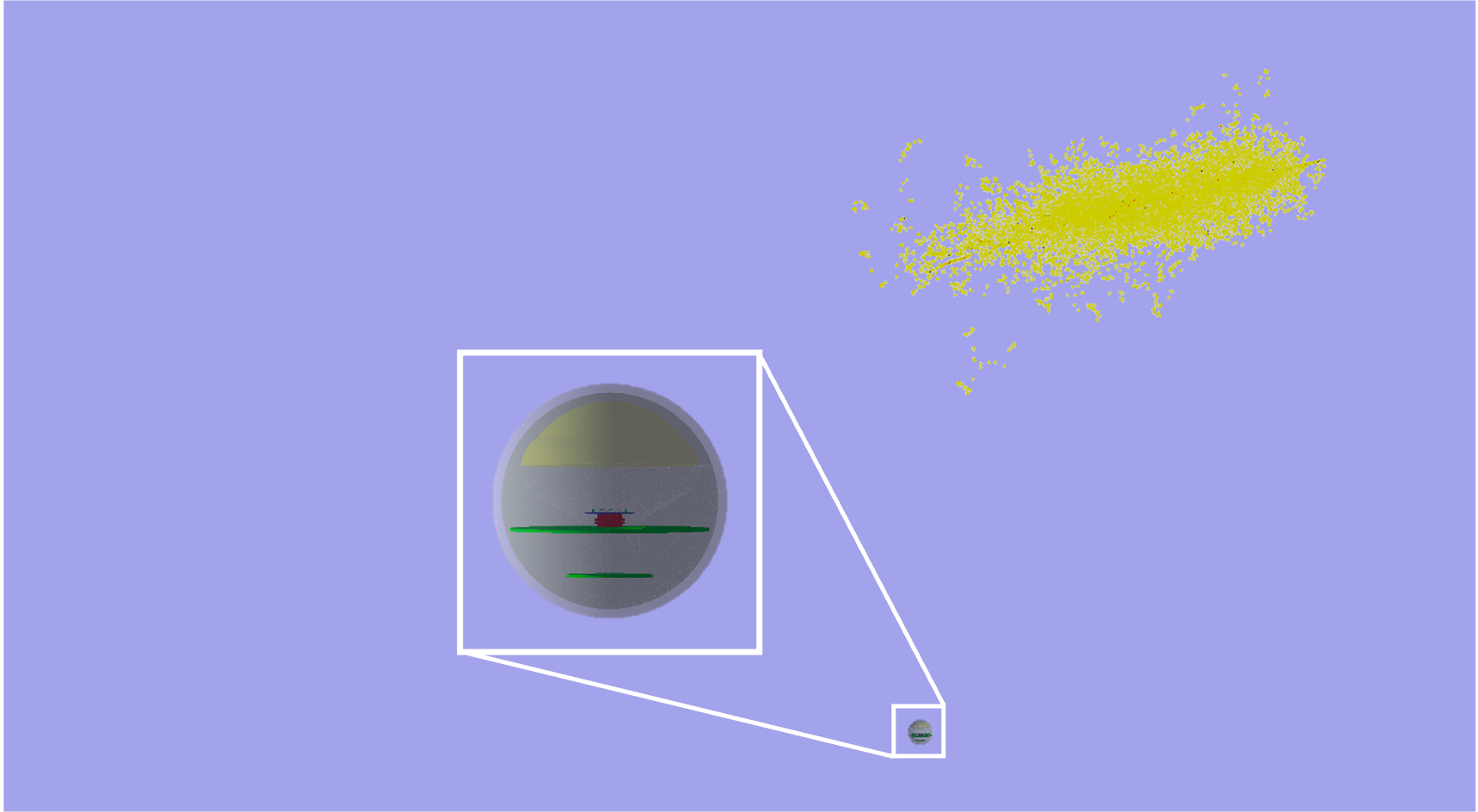}
  \caption{
  \Geant\ \texttt{OMSim} event display of a 50~GeV muon (left) and electron (right) passing by a DOM. The white rectangle zooms in on the DOM.
  Yellow points indicate the sampled trajectory steps of the charged particles (injected lepton and secondaries), Cherenkov photons are not shown.}
  \label{fig:Geant4_vis}
\end{figure}

Using \Geant\ and the IceCube \texttt{OMSim} framework~\cite{OMSim}, we simulate photon hits on a single DOM in ice according to the SPICE model~\cite{IceTransparency2013}. Primary muons and electrons are injected at 5–100~GeV and results are shown for $10$, $50$, and $80~\text{GeV}$, representative of the full range. The energy scan is used to examine how the particle-dependent timing
signature evolves across the energy range of interest, while keeping the
source--DOM geometry controlled through the impact parameter. \Geant\ propagates each particle and its secondaries, generates Cherenkov photons, and transports them through ice including scattering and absorption.
Fig.~\ref{fig:Geant4_vis} shows \Geant\ renderings of $50~\text{GeV}$ electrons and muons injected into ice in the single-DOM setup, including the secondary particles they produce.
The different locations of charged particles shown in \cref{fig:Geant4_vis} for muons and electrons already provide qualitative hints for the different PATDs discussed in section~\ref{subsec:patd}. We also record how each photon was created to investigate the origin of the PATD shapes. 

%The simulations also record the arrival times of photons reaching the sensitive area of the DOM.

\subsection{Photon arrival time distributions}
\label{subsec:patd}
\begin{figure}[htbp]
  \centering
  \includegraphics[width=.87\linewidth]{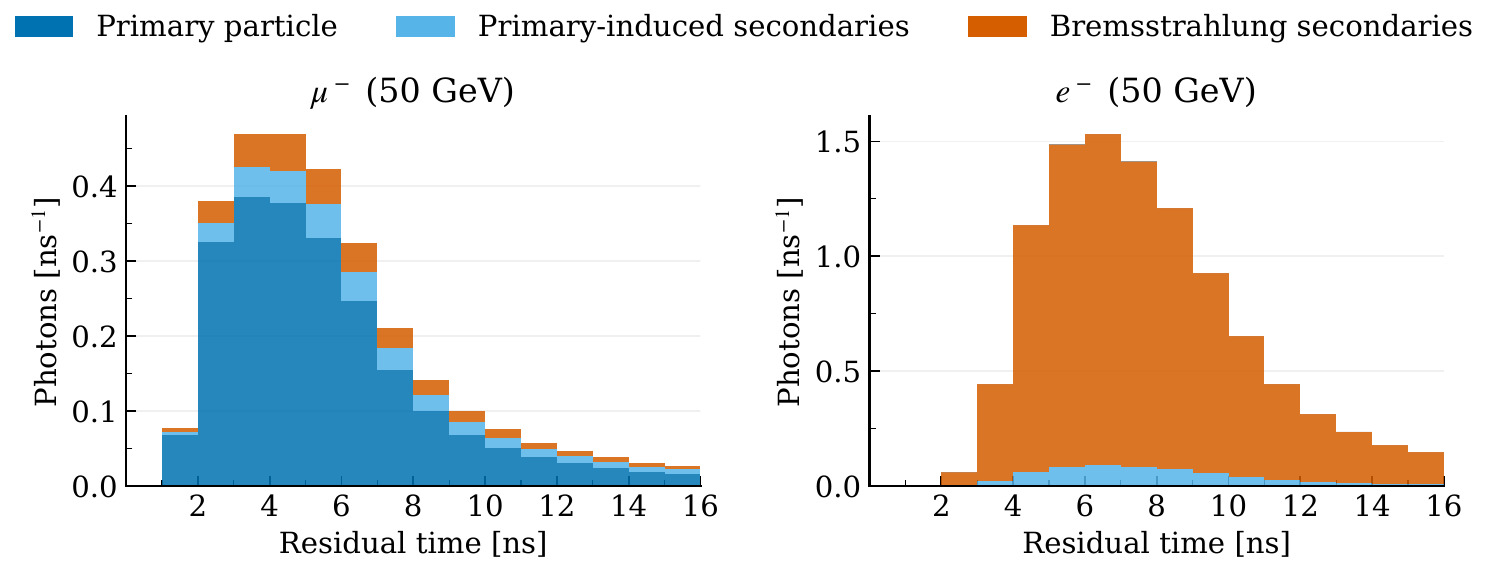}
  \caption{
  Average photon arrival time distribution for 50~GeV muons (left) and 
  electrons (right) at 20~m impact parameter. Muon events are dominated 
  by direct Cherenkov emission from the primary track, while electron 
  events are dominated by Bremsstrahlung secondaries.}
  \label{fig:patd_parent}
\end{figure}

The timing residual is defined as the difference between the observed photon 
arrival time and the expectation for unscattered Cherenkov light given the 
source–DOM geometry, and forms the x-axis of \cref{fig:patd_parent}. 
The muon PATD develops faster and is concentrated at early residual times, 
while the cascade PATD is broader. This difference follows from the 
light-production mechanism. Muon light is dominated by direct Cherenkov 
emission, whereas cascade light is dominated by Bremsstrahlung secondaries 
that are produced as the electromagnetic shower develops, spreading out the 
arrival times. The decompositions by photon origin in \cref{fig:patd_parent} make this explicit. The \emph{primary particle} label denotes direct Cherenkov light from the injected particle. \emph{Primary-induced secondaries} denotes light from particles produced in its electromagnetic interactions, such as $\delta$ rays from ionization. \emph{Bremsstrahlung secondaries} denotes light from $e^+e^-$ pairs produced by Bremsstrahlung photons radiated in the shower.

The absolute per-DOM photon yield and the spatial extent of the light
emission differ between the two particle types because of their
different propagation and shower geometries. A 50~GeV muon emits Cherenkov
light over more than 200~m of ice, whereas a 50~GeV electron-induced shower
is comparatively compact. When the light pattern is detected by sufficiently many DOMs, such spatial differences provide strong morphology-based PID information and are already exploited by existing reconstruction methods. Because the absolute photon yields differ between the two particle types, a normalized comparison of the PATD shapes, independent of the overall photon yield, is provided in fig.~\ref{fig:patd_pdf}.

The timing structure studied here is intended as a complementary information
channel. At the detector level, \WavePID\ targets events for which the event
morphology is only weakly constrained and exploits the relative temporal
distribution of the light recorded on individual DOMs. Across the
5--100~GeV range, muons behave as minimum ionizing particles (MIPs), and the muon and electron PATDs retain qualitatively different
early-time structure, motivating its use as an additional PID observable.

\begin{figure}[htbp]
  \centering
  \includegraphics[width=1\linewidth]{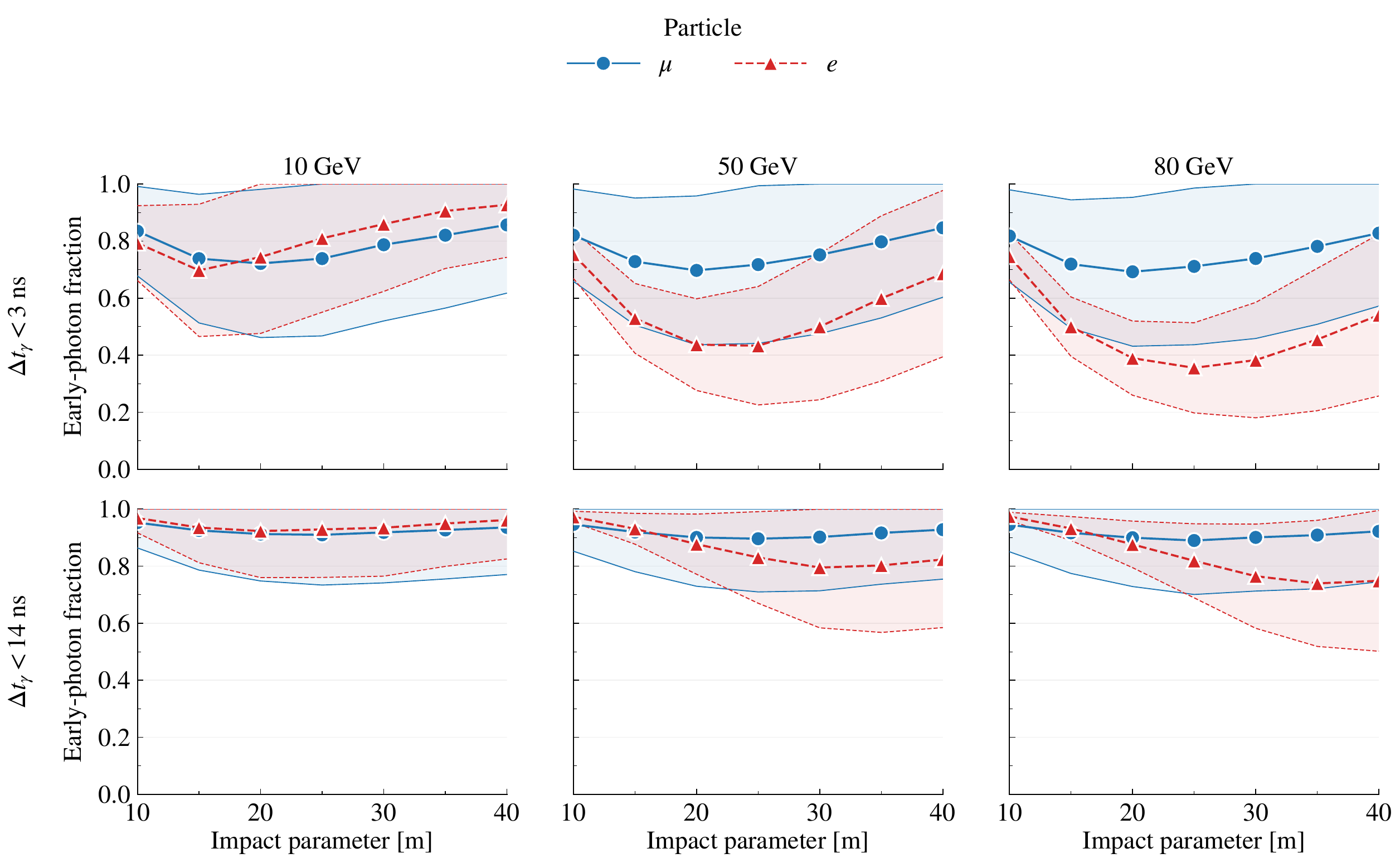}
\caption{Early-photon fraction (within $\Delta t$ of the first photon) as a 
function of impact parameter, for muons (blue) and electrons (red). Panels span three 
particle energies (10, 50, 80~GeV) and two integration windows 
($\Delta t = $ 3, 14~ns). Shaded regions indicate one standard deviation between events. Because the quantity is normalized to the total detected photon count,
differences in absolute photon yield are removed.
The 3~ns window illustrates the discrimination available at the photon 
level with high timing resolution, while the 14~ns window matches the 
timing adopted by \WavePID\ (cf. \cref{sec:wavepid}).}
  \label{fig:early_vs_distance}
\end{figure}

A useful derived observable is the \emph{early-photon fraction}, defined as
the number of photons arriving within a short window $\Delta t$ after the
first photon divided by the total number of detected photons. By
construction, this normalization removes the absolute photon yield and
isolates differences in the temporal distribution. Fig.~\ref{fig:early_vs_distance} shows this fraction as a 
function of the impact parameter for three energies and two integration 
windows. 

To isolate the microphysics origin of track--cascade differences, the \Geant\ study in this section evaluates early-photon fractions with integration windows as short as 3 ns, finer than the timing aggregation used on reconstructed IceCube pulses.
At the 3~ns window, the separation in ice reproduces that of the FNAL-1267 
water Cherenkov study~\cite{Samani2020}.
The 14~ns window provides a detector-relevant aggregation scale for comparison with the reconstructed-pulse implementation in \cref{sec:wavepid}. 
At 50 and 80~GeV with the 3~ns window, the separation is robust across 
impact parameters and peaks at 15--30~m. Broadening the time window to 14~ns reduces 
the separation at impact parameters below $\sim$20~m, where the characteristic 
Cherenkov peak from tracks is no longer resolved. At 10~GeV 
the early-photon fraction loses its separating power at the 14 ns window.

These photon-level trends show how the early-time signal evolves across the relevant energy range and motivate its use as a PID observable. At the detector level, such timing information may be particularly useful for events with a low hit multiplicity, where the event morphology provides limited discrimination. \WavePID\ combines the early-charge fraction with the DOM-to-vertex distance and the inter-module time difference in a multivariate template (\cref{sec:wavepid}). Its performance is evaluated with the full IceCube simulation in \cref{subsec:TS}, focusing on the $10 \leq n_{\rm hit} < 20$ regime.

\section{The \WavePID\ algorithm}
\label{sec:wavepid}

This section presents the \WavePID\ algorithm and its application to IceCube. The objective is to test whether the early-time signal identified in \cref{sec:Geant4} provides classification power beyond existing methods. To this end, \WavePID\ is constructed as a compact log-likelihood ratio over three physics-motivated observables, defined in \cref{tab:observables}. If this approach yields discrimination that a state-of-the-art GNN does not capture, the information content of the timing channel is indicated. While \WavePID\ is constructed without explicit dependence on hit multiplicity and can be applied across a broad energy range, this study focuses on the low-hit regime ($n_{\mathrm{hit}} < 20$) where morphology-based classification is most limited and timing-based approaches offer the largest relative improvement.

\paragraph{\WavePID's per-DOM observables.}
\WavePID\ operates on nanosecond-resolution ATWD pulses from individual DOMs, represented as an ordered set of reconstructed pulses $\{(t_p, q_p)\}$ with timestamps $t_p$ and charges $q_p$. Only pulses digitized by the ATWD are considered. For each DOM $n$, the algorithm constructs a feature vector $\mathbf{x}_n$ from the three observables defined in \cref{tab:observables}:
\begin{equation}
\mathbf{x}_n = \left(r_n,\ \xi_n^{t},\ \Delta t_n\right).
\end{equation}

\begin{table}[htbp]
\centering
\caption{Per-DOM observables used by \WavePID.}
\label{tab:observables}
\begin{tabular}{@{}lp{8cm}@{}}
\toprule
Symbol & Description \\
\midrule
$r_n$ & Distance from DOM $n$ to reconstructed interaction vertex; captures distance-dependent timing structure \\
$\xi_n^{t}$ & Fraction of charge on DOM $n$ arriving within first $t$\,ns; sensitive to early-time Cherenkov emission \\
$\Delta t_n$ & Time of first pulse on DOM $n$ relative to the earliest pulse on any DOM; encodes event-wide timing \\
\bottomrule
\end{tabular}
\end{table}
\paragraph{Template construction.}
The feature space $(r, \xi^t, \Delta t)$ is discretized into three-dimensional bins $\{\mathcal{B}_{ijk}\}$. For an event $e$, the occupancy tensor
\begin{equation}
H^e_{ijk} = \sum_{n \in e} \mathbb{I}\!\left[\mathbf{x}_n \in \mathcal{B}_{ijk}\right] ,
\end{equation}
counts the number of DOMs whose features fall into bin $\mathcal{B}_{ijk}$. 
This results in a 3D histogram of the DOMs in the feature space.
Shape-only templates $\lambda^s_{ijk}$ are constructed from labeled MC simulations for each hypothesis $s \in \{\mathrm{cascade}, \mathrm{track}\}$. Let $\mathcal{E}_s$ denote the set of MC events with true label $s$. 
Weighted bin counts are accumulated as weighted 3D histograms
\begin{equation}
N^s_{ijk} = \sum_{e \in \mathcal{E}_s} w^e \cdot H^e_{ijk},
\qquad
W^s = \sum_{e \in \mathcal{E}_s} w^e,
\end{equation}
where $w^e$ is the per-event weight~\cite{Andreopoulos:2009rq}, inherited by each DOM,
$N^s_{ijk}$ is the weighted DOM count in bin $\mathcal{B}_{ijk}$ for class $s$,
and $W^s$ is the total weight of class $s$. To ensure numerical stability, a small floor $\varepsilon_\lambda > 0$ is applied before normalizing to obtain proper shape-only probability mass functions:
\begin{equation}
\lambda^s_{ijk} = \frac{\tilde{\lambda}^s_{ijk}}{\sum_{i'j'k'} \tilde{\lambda}^s_{i'j'k'}}, \quad \text{with} \quad \tilde{\lambda}^s_{ijk} = \max\!\left(\varepsilon_\lambda,\, \frac{N^s_{ijk}}{W^s}\right).
\label{eq:lambda_main}
\end{equation}
The floor is set to $\varepsilon_\lambda = 10^{-12}$. Note that $\sum_{ijk} \tilde{\lambda}^s_{ijk}$ gives the average number of hits in an event of class $s$ up to the effect of this floor.

\paragraph{Test statistic.}
To discriminate between track and cascade morphologies, \WavePID\ employs a binned log-likelihood ratio. Assuming independent Poisson-distributed bin occupancies with expected bin occupancy $\mu^s_{ijk}$ under hypothesis $s$, the likelihood for hypothesis $s$ for an event $e$ is

\begin{equation}
\mathcal{L}_s(e) = \prod_{i,j,k} \mathrm{Pois}\!\left(H^e_{ijk} \mid \mu^s_{ijk}\right)
=
\prod_{i,j,k}
\frac{e^{-\mu^s_{ijk}}\left(\mu^s_{ijk}\right)^{H^e_{ijk}}}{H^e_{ijk}!}\,,
\label{eq:pois_product}
\end{equation}
where $\mu^s_{ijk} \propto \lambda^s_{ijk}$ and $\mu^s_{ijk}$ is normalized to the number of hits in an event $e$, i.e., $\sum_{ijk}\mu^s_{ijk}=\sum_{ijk}H^e_{ijk}$. The test statistic (TS) is defined as the shape-only log-likelihood ratio
\begin{equation}
\mathrm{TS}_e = \log\frac{\mathcal{L}_e^{\mathrm{cascade}}}{\mathcal{L}_e^{\mathrm{track}}}
= \sum_{i,j,k} H^e_{ijk} \log\frac{\lambda_{ijk}^{\mathrm{cascade}}}{\lambda_{ijk}^{\mathrm{track}}},
\label{eq:ts}
\end{equation}
representing a sum over bins of the per-bin log-odds, each weighted by the observed hits $H^e_{ijk}$. By construction, larger values of $\mathrm{TS}$ correspond to more cascade-like events. As the 
test statistic reduces to a lookup and summation over DOM feature bins, \WavePID\ is computationally inexpensive.

\section{\WavePID\ in IceCube}
\label{sec:results}
This section evaluates \WavePID\ performance on a set of 31 IceCube MC 
ensembles that represent the dominant detector and ice-model uncertainties, 
including DOM efficiency, the optical properties of the refrozen ice in the 
drill holes immediately surrounding each string (hole ice), and the 
scattering and absorption of the surrounding glacial ice 
(bulk ice)~\cite{IceTransparency2013}.

\subsection{Pulse timing on individual IceCube DOMs}
\begin{figure}[htbp]
    \centering
    \includegraphics[width=0.65\linewidth]{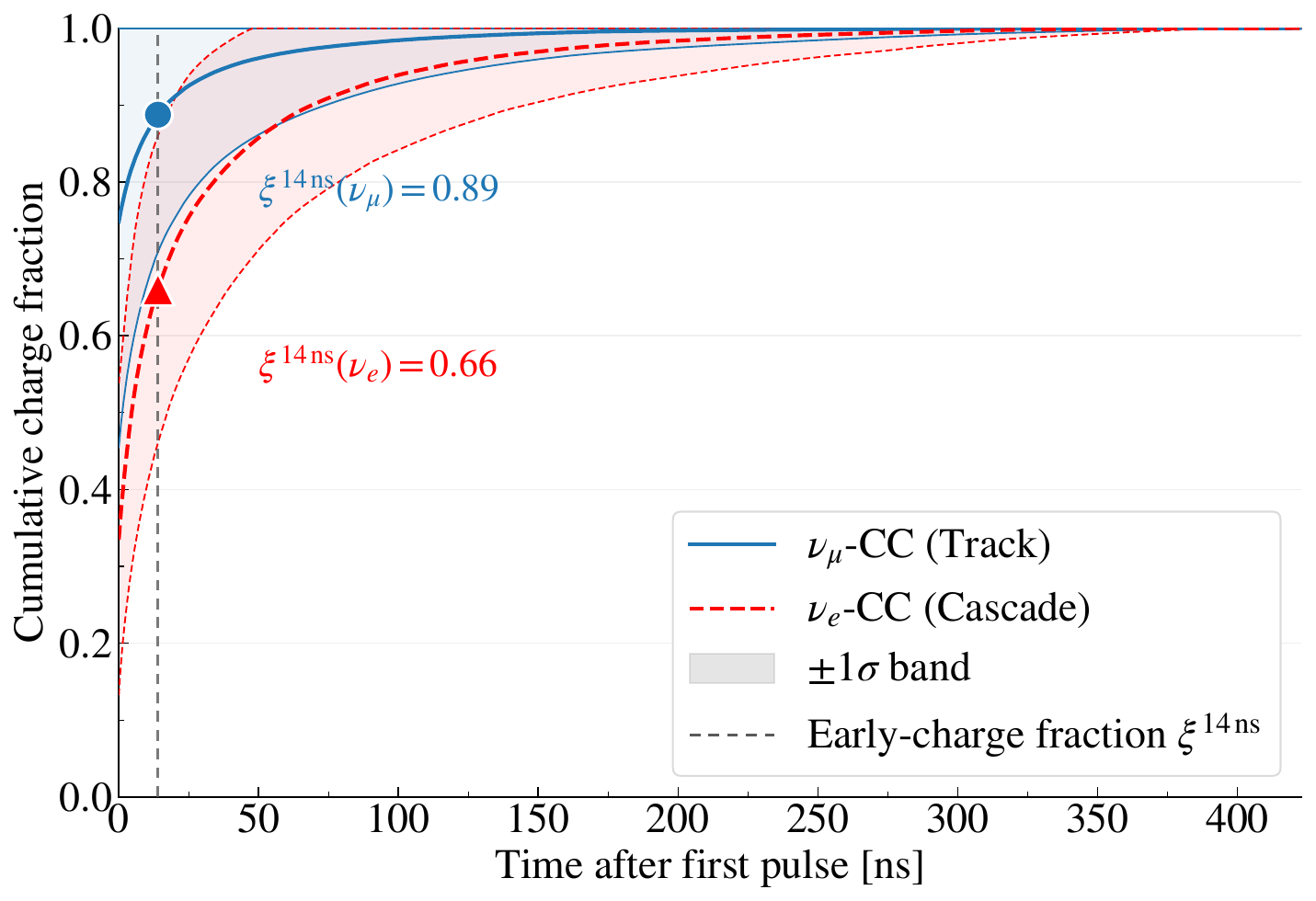}
    \caption{Cumulative charge fraction distribution for ${\nu}_\mu$ and ${\nu}_e$ charged-current interactions at 50~GeV neutrino energy with 35~GeV deposited by the charged lepton and 15~GeV by hadrons. Results are shown for a single DeepCore DOM and one representative event geometry. Shaded regions indicate one standard deviation across $10^5$ repetitions of the photon propagation, PMT response, and detector processing. The vertical dashed line marks 14~ns, which is defined as the early-charge fraction time window in \cref{subsec:TS}.}
    \label{fig:cdf_charge_frac}
\end{figure}

To examine whether the timing differences identified in the \Geant\ study 
persist through the full detector simulation chain, we use one representative 50~GeV $\nu_\mu$ event and one 
representative 50~GeV $\nu_e$ event. These events are chosen to 
represent typical deep-inelastic-scattering interactions, which dominate 
the studied energy range. The interaction vertex, directions, and the lepton and hadronic energy deposition are all held fixed. Starting from this fixed simulated event,
we independently resample the stochastic photon propagation and PMT response
$10^5$ times. For each realization, the resulting photon hits are converted to pulses using the standard pulse extraction and passed through the DeepCore pulse cleaning and filtering used by \WavePID. No event-level reconstruction is performed for these resampled realizations, and the DOM-to-vertex distance is not used here. This procedure isolates fluctuations from photon propagation and detector response for a fixed underlying event.
The full per-event complexity is present in the \WavePID\ evaluation on the complete MC sample, presented in \cref{subsec:TS}.
Fig.~\ref{fig:cdf_charge_frac} shows that the early-charge fraction retains 
discriminating power in this geometry after the complete simulation chain, 
indicating that the signal survives the detector response.

This motivates compiling the per-DOM observables into the binned templates described in \cref{sec:wavepid}.
The \WavePID\ test statistic is then obtained by 
summing the bin-wise log-odds over all contributing DOMs in an event, as evaluated below.
\subsection{Test statistic performance and systematics}
\label{subsec:TS}
We present the implementation and performance of \WavePID\ on IceCube DeepCore.
The study uses the oscNext event selection~\cite{DynEdge2022, Kozynets:2024phd}, which
provides a high-purity selection of neutrino events in the 5--100~GeV range with atmospheric muon contamination below 1\%. The MC-truth particle and interaction type are used for the construction of the TS templates, where $\nu_\mu$ CC interactions are defined as tracks and all other neutrino interactions as cascades. The simulation used here relies on GENIE v2.12.8~\cite{Andreopoulos:2009rq}, which does not include Meson Exchange Current (MEC). MEC will mainly impact the amount of energy given to the outgoing lepton and hadron, which can in principle change the expected light yield of low energy tracks and cascades. We leave detailed evaluation of the impact of MEC on \WavePID\ to future work. 
We restrict the sample to the cascade-PID bin of the oscNext selection, defined by a \DynEdge-PID score $\leq 0.4$, where lower values correspond to more cascade-like events. We further require $10 \leq n_\mathrm{hit} < 20$, focusing on the target regime of this method where morphology-based classification is weakest.
%after atmospheric muon rejection and DeepCore containment cuts. 
%The dataset comprises $1.19 \times 10^5$ events over 11.1~years, with atmospheric muon contamination at the $\sim$1\% level. 
%For template construction, tracks are defined as $\nu_\mu$~CC interactions and cascades as all other neutrino interactions.
\paragraph{Reconstruction and binning.}
For the full-event \WavePID\ evaluation, the DOM-to-vertex distance $r$ is
computed for each DOM using the interaction vertex reconstructed by \DynEdge\
for that event. The bin edges used in this study
are
\begin{equation}
\begin{aligned}
\mathbf{e}_r &= (0,\ 35,\ 50,\ 65,\ 84,\ 111,\ 1500)\ \mathrm{m},\\
\mathbf{e}_\xi &= (0.0,\ 0.1,\ 0.2,\ \ldots,\ 0.9,\ 1.0),\\
\mathbf{e}_{\Delta t} &= (0,\ 5,\ 36,\ 253,\ 1800)\ \mathrm{ns},
\end{aligned}
\label{eq:binning}
\end{equation}
where the distance bins in $\mathbf{e}_r$ are defined by quantiles of
the distance distribution over the full MC sample, the early-charge
fraction~$\xi^t$ is binned uniformly on $[0,1]$ for $t = 14~\mathrm{ns}$,
and the $\Delta t$ bins are log-spaced from $5$ to $1800~\mathrm{ns}$,
with a separate first bin $[0,5)~\mathrm{ns}$ for the earliest pulse
arrivals.
For the selected event sample, the median three-dimensional distance between
the \DynEdge-reconstructed and true vertices is $8.6~\mathrm{m}$, with
$7.8~\mathrm{m}$ for cascades and $13.7~\mathrm{m}$ for tracks. This
reconstruction uncertainty dominates the uncertainty on~$r$, exceeding the $\sim$1~m uncertainty on DOM
positions~\cite{Abbasi:20231J}. A dedicated check confirms that adding a Gaussian smearing of $\sigma = 10$~m 
per axis to the reconstructed vertex leaves the area under the receiver-operating-characteristic (ROC) curve (AUC) stable to within $\sim 0.02$, consistent with the 
chosen distance binning being coarse relative to the \DynEdge\ vertex 
resolution.

The 14~ns time window is large compared to both the PMT transit time spread
($\sim$3.2~ns~\cite{Abbasi2010vc}) and the timing accuracy of the
Reciprocal Active Pulsing (RapCal) calibration system
($\sim$1.2~ns~\cite{IceCube2017JINST}), indicating that instrumental timing
uncertainties are subdominant to the physical timing differences exploited
by the early-charge fraction.
The first $[0,5)~\mathrm{ns}$ bin of $\Delta t$ groups DOMs with the
earliest first-pulse times in an event and isolates the prompt-arrival
region of the likelihood template. This bin should not be interpreted as a
$5~\mathrm{ns}$ instrumental timing resolution or as a distinct physical
timescale. The PMT transit-time response is included in the nominal detector
simulation used to construct the $\Delta t$ templates. The residual inter-DOM
timing uncertainty after RapCal calibration is approximately
$1.2~\mathrm{ns}$, smaller than the $[0,5)~\mathrm{ns}$ bin width. This
timing calibration uncertainty is not varied independently in the
detector-systematic ensembles considered here.

The bin edges and time window were optimized empirically
for PID performance in events with 10--20 hits, informed by the
distance and timing dependencies identified in
section~\ref{sec:Geant4}. Hyperparameters were fixed on the training
set prior to evaluation on the test set.

\paragraph{Performance and benchmarking.}
The cascade and track templates for \WavePID\ are constructed from an 
independent 80\% training set and all \WavePID\ and \DynEdge\ ROC curves 
are evaluated on the held-out 20\% test set. Both classifiers are 
benchmarked within the \DynEdge-selected cascade sample, so the reported AUCs measure 
residual flavor-separation power after the GNN pre-selection.

\begin{figure}[htbp]
  \centering
  \includegraphics[width=0.85\linewidth]{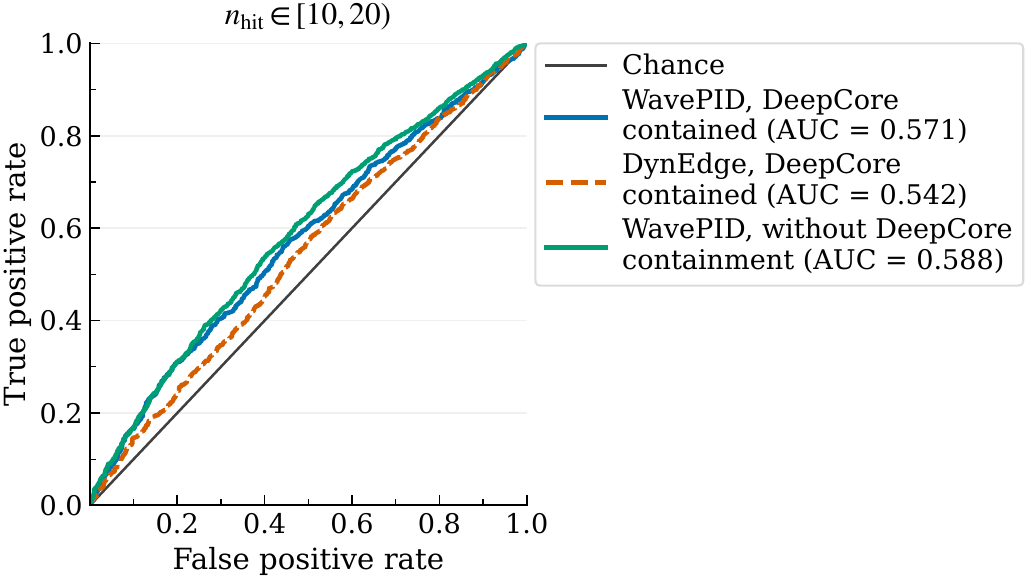}
  \caption{PID performance evaluated on a held-out 20\% test set, for cascade 
classification in the \DynEdge\ cascade PID bin with $n_\mathrm{hit} \in [10, 20)$. 
\WavePID\ (blue, solid) achieves higher AUC than \DynEdge\ (orange, dashed) 
within the standard DeepCore-contained event selection. Relaxing the DeepCore containment requirement (green, solid) slightly increases separation power, as \WavePID\ does not rely on the containment cut and benefits from the larger sample.}
  \label{fig:roc_benchmark}
\end{figure}

As shown in \cref{fig:roc_benchmark}, for $10 \leq n_\mathrm{hit} < 20$, 
\WavePID\ achieves higher AUC than \DynEdge\ (0.57 vs.\ 0.54) despite 
relying on only three observables, demonstrating that early-time pulse 
structure encodes PID information not captured by the GNN. Both ROC curves 
are evaluated within the \DynEdge\ cascade PID bin and sweep each classifier's 
threshold across the score range available in that sub-sample, so they span 
the full $[0,1]$ true positive rate and false positive rate range by construction. The \DynEdge\ AUC 
reflects the residual ordering power of the \DynEdge\ score within its own 
cascade bin, with most of \DynEdge's separating power having been used by 
the upstream cut at \DynEdge-PID $\leq 0.4$.
For larger hit multiplicities, \DynEdge\ achieves higher AUC, consistent 
with its ability to exploit richer topological information, confirming that 
the two methods access complementary features. The AUC margin itself is 
secondary to this complementarity, positioning \WavePID\ as an additional 
classification dimension rather than a standalone alternative to 
morphology-based methods.

\begin{figure}[htbp]
  \centering  \includegraphics[width=0.95\linewidth]{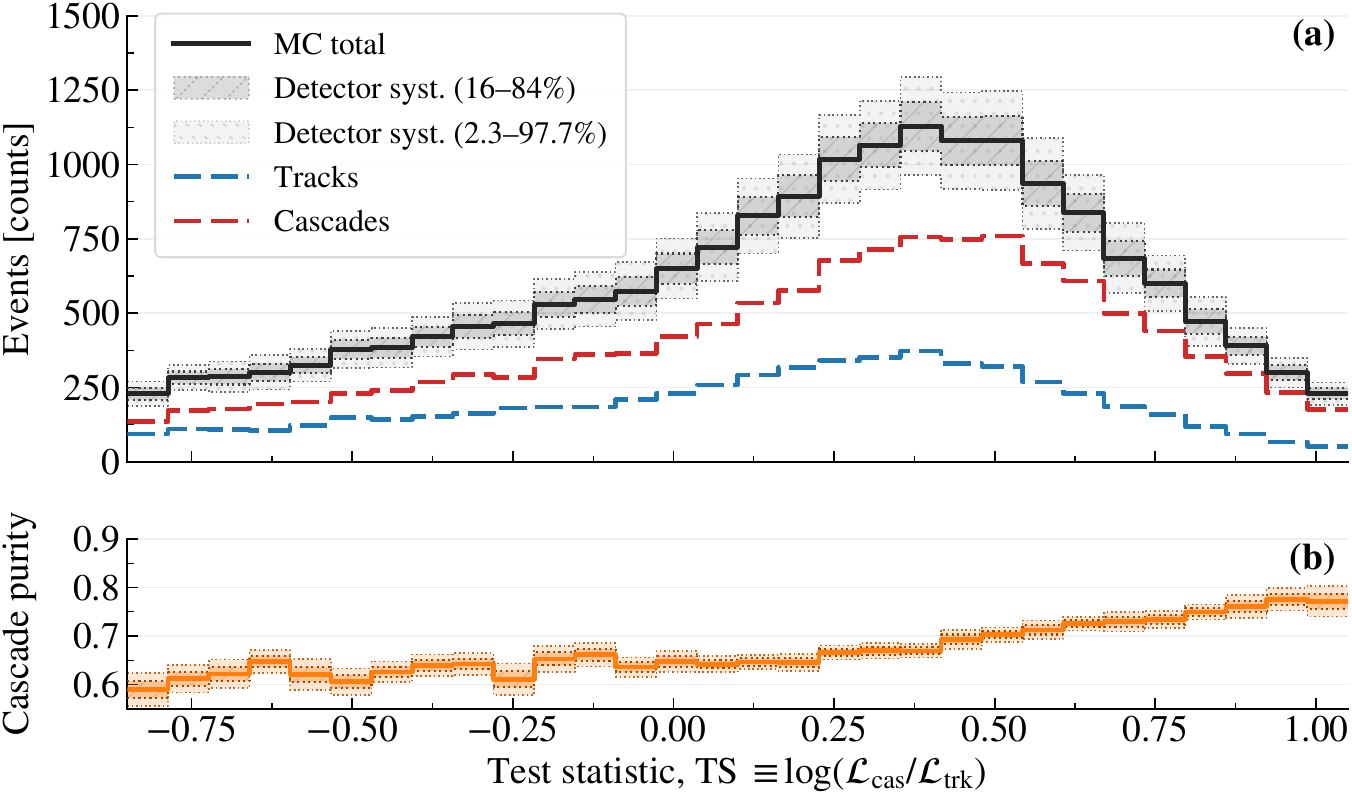}
  \caption{Top panel~(a): \WavePID\ test statistic distribution in the \DynEdge\ cascade PID bin for simulated true tracks ($\nu_\mu$ CC) and cascades (all other neutrino interactions), computed as the shape-only log-likelihood ratio. The shaded bands indicate the 16--84th and the 2.3--97.7th percentile variation across 31 detector-systematics MC sets. Lower panel~(b): cascade purity.}\label{fig:ts_distribution}
\end{figure}

\paragraph{Systematics.}
The \WavePID\ template is constructed only once from the nominal MC set
and evaluated on all 31 oscNext detector-systematic ensembles, which sample
DOM efficiency, hole-ice parameters, and bulk-ice optical properties.
The PMT transit-time spread is included in the nominal detector simulation
through the standard PMT-response simulation, but its uncertainty is not
varied independently in these MC ensembles. Residual uncertainties in the
inter-DOM timing synchronization are likewise not varied as an independent
detector systematic. These uncertainties are therefore not represented by
the systematic bands shown here. The resulting spread is shown as shaded
bands in \cref{fig:ts_distribution}, where the TS shape and track--cascade
separation remain stable across all configurations. A full treatment of
physics-parameter uncertainties and additional detector effects is left for
dedicated physics analyses, as discussed in \cref{sec:conclusion}.

\paragraph{Purity increase.}
Applying a threshold $\mathrm{TS} > \mathrm{TS}_{\mathrm{cut}}$ to the \WavePID\ 
output increases the cascade purity $P^{\mathrm{C}}$ within the \DynEdge\ cascade 
selection. At a representative operating point retaining 20\% of cascades, the 
baseline purity from \DynEdge\ of $P^{\mathrm{C}}_{0} = 0.66$ increases to 
$P^{\mathrm{C}} = 0.75$ for the nominal MC set. Beyond the nominal set, across the 
31 detector-systematic ensembles, the purity gain is $+7\%\pm 1\%$, indicating 
robustness to the dominant detector uncertainties. The 20\% operating point is 
illustrative rather than prescriptive. A hard cut at this operating point keeps 
only whether each event passes the threshold and discards the continuous per-event 
score. In a physics analysis the test statistic or the early-charge fraction is more useful as an additional input variable. Quantification of the resulting physics impact is left to dedicated follow-up studies
(\cref{sec:conclusion}).

%------------------------------------------------------------------------------
\section{Conclusion and outlook}
\label{sec:conclusion}
We presented \WavePID, a compact template-based classifier that exploits 
nanosecond-scale pulse timing for flavor identification in IceCube DeepCore 
at 5--100~GeV. We demonstrate nanosecond-scale Cherenkov timing at individual DOMs as a previously unexploited, first-principles-motivated information channel for flavor identification in neutrino telescopes. 

Waveform-based PID 
with single PMTs has previously been demonstrated in controlled beam 
measurements~\cite{Samani2020}, and here we show that the same timing 
structure remains effective after propagation through South Pole ice and 
digitization by IceCube DOMs, with discrimination concentrated in the regime 
where the primary lepton's leading, minimally scattered Cherenkov front is 
resolved by per-DOM pulse timing.
Using only three physics-motivated observables (DOM-to-vertex distance, 
early-charge fraction, and inter-module time difference), \WavePID\ 
achieves higher AUC than \DynEdge\ within the \DynEdge-selected cascade 
sample in the low-hit regime ($10 \leq n_\mathrm{hit} < 20$).
A microphysics interpretation is provided by 
\Geant\ photon arrival-time studies, which associate the observed 
discrimination with differences in Cherenkov emission geometry between muon 
tracks and electromagnetic showers. These results show that early 
nanosecond-scale timing carries discriminating information beyond that 
captured by the existing event-level classifier in this regime, with \WavePID's three physics-motivated observables complementary to the learned representation that \DynEdge\ builds from the raw pulses.

The configuration presented here defines a baseline implementation, and the 
optimal choice of observables, binning, and operating point depends on the 
physics target. In practice, the natural application of \WavePID\ is as an 
additional classification dimension, where the test statistic, or the 
early-charge fraction alone, can be combined with the score of a classifier like \DynEdge-PID\ in a 
multi-dimensional selection or used as an input to a likelihood fit, retaining 
the full event sample while exploiting per-event timing information.

Application to IceCube data and quantification of the corresponding physics 
impact are left to dedicated follow-up studies. Extensions such as 
electromagnetic versus hadronic cascade separation are natural next steps. 
More broadly, \WavePID\ motivates the development of fine-grained per-module 
timing in next-generation neutrino telescopes operating in ice or water, 
such as the IceCube Upgrade~\cite{IceCubeUpgrade_mDOM_ICRC2021} and 
IceCube-Gen2~\cite{IceCubeGen2_2021}, 
KM3NeT/ORCA~\cite{KM3NeT_multiPMT_2022, KM3NeT_ORCA_NMO_2022}, 
P-ONE~\cite{POne_2020}, TRIDENT~\cite{TRIDENT_NatAstron_2023}, and 
HUNT~\cite{HUNT_ICRC2023}, particularly for detectors with multi-PMT or hybrid 
optical modules where per-module timing is optimized by design.

%------------------------------------------------------------------------------
% \acknowledgments
%We acknowledge TBD.
\acknowledgments
The IceCube collaboration acknowledges the significant contributions to this manuscript from Steven Young Eulig. 
The authors gratefully acknowledge the support from the following agencies and institutions:
USA {\textendash} U.S. National Science Foundation-Office of Polar Programs,
U.S. National Science Foundation-Physics Division,
U.S. National Science Foundation-EPSCoR,
U.S. National Science Foundation-Office of Advanced Cyberinfrastructure,
Wisconsin Alumni Research Foundation,
Center for High Throughput Computing (CHTC) at the University of Wisconsin{\textendash}Madison,
Open Science Grid (OSG),
Partnership to Advance Throughput Computing (PATh),
Advanced Cyberinfrastructure Coordination Ecosystem: Services {\&} Support (ACCESS),
Frontera and Ranch computing project at the Texas Advanced Computing Center,
U.S. Department of Energy-National Energy Research Scientific Computing Center,
Particle astrophysics research computing center at the University of Maryland,
Michigan State University,
Astroparticle physics computational facility at Marquette University,
NVIDIA Corporation,
and Google Cloud Platform;
Belgium {\textendash} Funds for Scientific Research (FRS-FNRS and FWO),
FWO Odysseus and Big Science programmes,
and Belgian Federal Science Policy Office (Belspo);
Germany {\textendash} Bundesministerium f{\"u}r Forschung, Technologie und Raumfahrt (BMFTR),
Deutsche Forschungsgemeinschaft (DFG),
Helmholtz Alliance for Astroparticle Physics (HAP),
Initiative and Networking Fund of the Helmholtz Association,
Deutsches Elektronen Synchrotron (DESY),
and High Performance Computing cluster of the RWTH Aachen;
Sweden {\textendash} Swedish Research Council,
Swedish Polar Research Secretariat,
Swedish National Infrastructure for Computing (SNIC),
and Knut and Alice Wallenberg Foundation;
European Union {\textendash} EGI Advanced Computing for research;
Australia {\textendash} Australian Research Council;
Canada {\textendash} Natural Sciences and Engineering Research Council of Canada,
Calcul Qu{\'e}bec, Compute Ontario, Canada Foundation for Innovation, WestGrid, and Digital Research Alliance of Canada;
Denmark {\textendash} Villum Fonden, Carlsberg Foundation, and European Commission;
New Zealand {\textendash} Marsden Fund;
Japan {\textendash} Japan Society for Promotion of Science (JSPS), Ministry of Education, Culture, Sports, Science and Technology (MEXT), and Institute for Global Prominent Research (IGPR) of Chiba University;
Korea {\textendash} National Research Foundation of Korea (NRF);
Switzerland {\textendash} Swiss National Science Foundation (SNSF).

\bibliographystyle{JHEP}
\bibliography{biblio}

@misc{OMSim,
  collaboration = {IceCube},
  title = {{OMSim}: {Geant4} framework for simulating optical modules of the {IceCube} observatory},
  year         = {2025},
  howpublished = {GitHub repository},
  url          = {https://github.com/icecube/OMSim},
  note         = {Accessed: 2025-12-23}
}

@article{Abbasi2010vc,

  author        = {Aartsen, M. G. and others},
  collaboration = {IceCube},
    title     = {Calibration and characterization of the {IceCube} photomultiplier tube},
    journal   = {Nucl. Instrum. Meth. A},
    volume    = {618},
    pages     = {139--152},
    year      = {2010},
    eprint    = {1002.2442},
    archivePrefix = {arXiv},
    primaryClass  = {astro-ph.IM},
    doi       = {10.1016/j.nima.2010.03.102}
}

@article{Abbasi:20231J,
  author        = {Aartsen, M. G. and others},
  collaboration = {IceCube},
  title = "{Refining the IceCube detector geometry using muon and LED calibration data}",
  doi = "10.22323/1.444.0988",
  journal = "PoS",
  year = 2023,
  volume = "ICRC2023",
  pages = "988"
}

@article{IceCube2017JINST,
  author        = {Aartsen, M. G. and others},
  collaboration = {IceCube},
title = {The {IceCube} neutrino observatory: Instrumentation and online systems},

  journal       = {JINST},
  volume        = {12},
  number        = {03},
  pages         = {P03012},
  year          = {2017},
  doi           = {10.1088/1748-0221/12/03/P03012},
  eprint        = {1612.05093},
  archivePrefix = {arXiv},
  primaryClass  = {astro-ph.IM}
}

@article{IceTransparency2013,
  author        = {Aartsen, M. G. and others},
  collaboration = {IceCube},
  title = {Measurement of {South Pole} ice transparency with the {IceCube} {LED} calibration system},
  journal       = {Nucl. Instrum. Meth. A},
  volume        = {711},
  pages         = {73--89},
  year          = {2013},
  doi           = {10.1016/j.nima.2013.01.054},
  eprint        = {1301.5361},
  archivePrefix = {arXiv},
  primaryClass  = {astro-ph.IM}
}

@article{LowEnergyReco2022,
  author        = {Abbasi, R. and others},
  collaboration = {IceCube},
  title         = {Low energy event reconstruction in {IceCube} {DeepCore}},
  journal       = {Eur. Phys. J. C},
  volume        = {82},
  pages         = {807},
  year          = {2022},
  doi           = {10.1140/epjc/s10052-022-10721-2},
  eprint        = {2203.02303},
  archivePrefix = {arXiv},
  primaryClass  = {physics.ins-det}
}

@article{DeepCoreOsc2023,
  author        = {Abbasi, R. and others},
  collaboration = {IceCube},
title = {Measurement of atmospheric neutrino mixing with improved {IceCube} {DeepCore} calibration and data processing},

  journal       = {Phys. Rev. D},
  volume        = {108},
  number        = {1},
  pages         = {012014},
  year          = {2023},
  doi           = {10.1103/PhysRevD.108.012014},
  eprint        = {2304.12236},
  archivePrefix = {arXiv},
  primaryClass  = {hep-ex}
}

@article{DeepCoreMassOrdering2020,
  author        = {Abbasi, R. and others},
  collaboration = {IceCube},
title = {Development of an analysis to probe the neutrino mass ordering with atmospheric neutrinos using three years of {IceCube} {DeepCore} data},

  journal       = {Eur. Phys. J. C},
  volume        = {80},
  number        = {1},
  pages         = {9},
  year          = {2020},
  doi           = {10.1140/epjc/s10052-019-7555-0},
  eprint        = {1902.07771},
  archivePrefix = {arXiv},
  primaryClass  = {hep-ex}
}

@article{DeepCoreSterile2024,
  author        = {Abbasi, R. and others},
  collaboration = {IceCube},
title = {Search for a light sterile neutrino with 7.5 years of {IceCube} {DeepCore} data},

  journal       = {Phys. Rev. D},
  volume        = {110},
  number        = {7},
  pages         = {072007},
  year          = {2024},
  doi           = {10.1103/PhysRevD.110.072007},
  eprint        = {2407.01314},
  archivePrefix = {arXiv},
  primaryClass  = {hep-ex}
}

@article{DynEdge2022,
  author        = {Abbasi, R. and others},
  collaboration = {IceCube},
title = {Graph neural networks for low-energy event classification and reconstruction in {IceCube}},
  journal       = {JINST},
  volume        = {17},
  number        = {11},
  pages         = {P11003},
  year          = {2022},
  doi           = {10.1088/1748-0221/17/11/P11003},
  eprint        = {2209.03042},
  archivePrefix = {arXiv},
  primaryClass  = {physics.ins-det}
}

@article{Samani2020,
  author        = {Samani, S. and others},
  title         = {Pulse shape particle identification by a single large hemispherical photomultiplier tube},
  journal       = {JINST},
  volume        = {15},
  number        = {05},
  pages         = {T05002},
  year          = {2020},
  doi           = {10.1088/1748-0221/15/05/T05002},
  eprint        = {1912.03901},
  archivePrefix = {arXiv},
  primaryClass  = {physics.ins-det}
}

@article{Geant42003,
  author  = {Agostinelli, S. and others},
  title   = {Geant4---a simulation toolkit},
  journal = {Nucl. Instrum. Meth. A},
  volume  = {506},
  pages   = {250--303},
  year    = {2003},
  doi     = {10.1016/S0168-9002(03)01368-8}
}

@phdthesis{Kozynets:2024phd,
  author  = {Kozynets, Tetiana},
  title   = {Atmospheric neutrino oscillations in IceCube-DeepCore within and beyond the unitary framework},
  school  = {University of Copenhagen, Faculty of Science},
  address = {Copenhagen, Denmark},
  year    = {2024},
  month   = nov,
  note    = {Ph.D. thesis, Niels Bohr Institute},
  url     = {https://nbi.ku.dk/english/theses/phd-theses/tetiana-kozynets/Tetiana-Kozynets.pdf}
}

@article{IceCube:2011ucd,
    author = "Abbasi, R. and others",
    collaboration = "IceCube",
    title = "{The Design and Performance of IceCube DeepCore}",
    eprint = "1109.6096",
    archivePrefix = "arXiv",
    primaryClass = "astro-ph.IM",
    doi = "10.1016/j.astropartphys.2012.01.004",
    journal = "Astropart. Phys.",
    volume = "35",
    pages = "615--624",
    year = "2012"
}

@article{IceCube:2019dqi,
    author = "Aartsen, M. G. and others",
    collaboration = "IceCube",
    title = "{Measurement of Atmospheric Tau Neutrino Appearance with IceCube DeepCore}",
    eprint = "1901.05366",
    archivePrefix = "arXiv",
    primaryClass = "hep-ex",
    doi = "10.1103/PhysRevD.99.032007",
    journal = "Phys. Rev. D",
    volume = "99",
    number = "3",
    pages = "032007",
    year = "2019"
}

@article{IceCube:2024nhk,
    author = "Abbasi, R. and others",
    collaboration = "IceCube",
    title = "{Measurement of Atmospheric Neutrino Oscillation Parameters Using Convolutional Neural Networks with 9.3 Years of Data in IceCube DeepCore}",
    eprint = "2405.02163",
    archivePrefix = "arXiv",
    primaryClass = "hep-ex",
    doi = "10.1103/PhysRevLett.134.091801",
    journal = "Phys. Rev. Lett.",
    volume = "134",
    number = "9",
    pages = "091801",
    year = "2025"
}

@article{Andreopoulos:2009rq,
    author = "Andreopoulos, C. and others",
    title = "{The GENIE Neutrino Monte Carlo Generator}",
    eprint = "0905.2517",
    archivePrefix = "arXiv",
    primaryClass = "hep-ph",
    doi = "10.1016/j.nima.2009.12.009",
    journal = "Nucl. Instrum. Meth. A",
    volume = "614",
    pages = "87--104",
    year = "2010"
}

@article{KM3NeT_multiPMT_2022,
    title = {The {KM3NeT} multi-{PMT} optical module},
    author = {{KM3NeT Collaboration}},
    journal = {JINST},
    volume = {17},
    pages = {P07038},
    year = {2022},
    eprint = {2203.10048},
    archivePrefix = {arXiv},
    primaryClass = {astro-ph.IM},
    doi = {10.1088/1748-0221/17/07/P07038},
}

@article{KM3NeT_ORCA_NMO_2022,
    title = {Determining the neutrino mass ordering and oscillation parameters with {KM3NeT/ORCA}},
    author = {{KM3NeT Collaboration}},
    journal = {Eur. Phys. J. C},
    volume = {82},
    pages = {26},
    year = {2022},
    eprint = {2103.09885},
    archivePrefix = {arXiv},
    primaryClass = {hep-ex},
    doi = {10.1140/epjc/s10052-021-09893-0},
}

@article{IceCubeGen2_2021,
    title = {{IceCube-Gen2}: the window to the extreme {U}niverse},
    author = {{IceCube-Gen2 Collaboration} and Aartsen, M. G. and others},
    journal = {J. Phys. G},
    volume = {48},
    pages = {060501},
    year = {2021},
    eprint = {2008.04323},
    archivePrefix = {arXiv},
    primaryClass = {astro-ph.HE},
    doi = {10.1088/1361-6471/abbd48},
}

@article{POne_2020,
    title = {The {P}acific {O}cean {N}eutrino {E}xperiment},
    author = {Agostini, M. and others},
    collaboration = {{P-ONE}},
    journal = {Nature Astron.},
    volume = {4},
    pages = {913--915},
    year = {2020},
    eprint = {2005.09493},
    archivePrefix = {arXiv},
    primaryClass = {astro-ph.HE},
    doi = {10.1038/s41550-020-1182-4},
}

@article{TRIDENT_NatAstron_2023,
    title = {A multi-cubic-kilometre neutrino telescope in the western {P}acific {O}cean},
    author = {Ye, Z. P. and others},
    journal = {Nature Astron.},
    volume = {7},
    pages = {1497--1505},
    year = {2023},
    eprint = {2207.04519},
    archivePrefix = {arXiv},
    primaryClass = {astro-ph.HE},
    doi = {10.1038/s41550-023-02087-6},
}

@article{HUNT_ICRC2023,
  author  = {Huang, Tian-Qi and Cao, Zhen and Chen, Mingjun and Liu, Jiali and Wang, Zike and You, Xiaohao and Qi, Ying},
  title   = "{Proposal for the High Energy Neutrino Telescope}",
  doi     = "10.22323/1.444.1080",
  journal = "PoS",
  year    = 2023,
  volume  = "ICRC2023",
  pages   = "1080"
}

@article{IceCubeUpgrade_mDOM_ICRC2021,
  author  = {{IceCube Collaboration}},
  title   = "{Design and performance of the multi-PMT optical module for IceCube Upgrade}",
  doi     = "10.22323/1.395.1070",
  journal = "PoS",
  year    = 2021,
  volume  = "ICRC2021",
  pages   = "1070",
  eprint  = "2107.11383",
  archivePrefix = "arXiv",
  primaryClass = "astro-ph.IM"
}

%\end{document}

\clearpage
\appendix
\setcounter{page}{1}
\renewcommand{\thepage}{S\arabic{page}}

\supplementpagestyle
\pagestyle{suppl}
%\end{document}
\section{Supplemental Material}
\paragraph{Additional \Geant\ figures.}
Supplementary event displays at 5 and 100~GeV illustrate that the qualitative differences between muon- and electron-induced event morphologies persist across the simulated energy range.

\begin{figure}[htbp]
  \centering
  \includegraphics[width=0.48\linewidth]{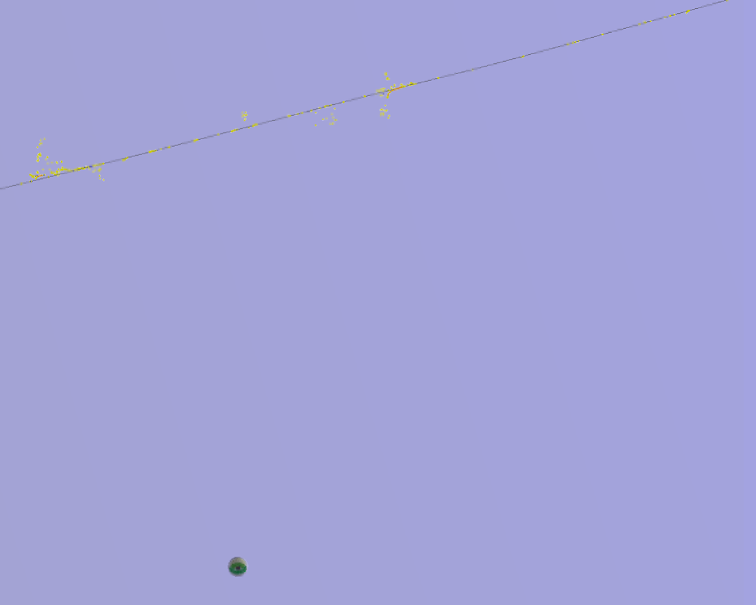}\hfill
  \includegraphics[width=0.48\linewidth]{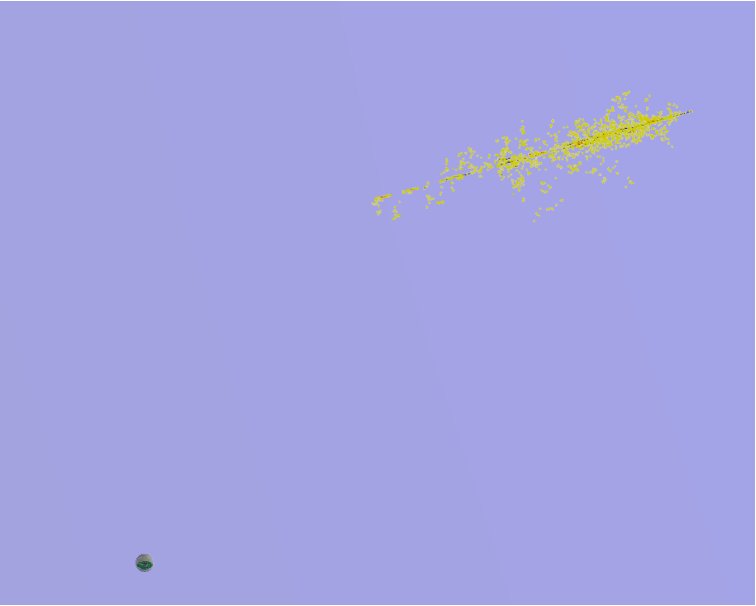}
  \caption{
  \Geant\ \texttt{OMSim} event display of a 5~GeV muon (left) and electron (right) near a DOM.
  Yellow points indicate the sampled trajectory steps of the charged particles (injected lepton and secondaries), Cherenkov photons are not shown.
  }
  \label{fig:geant4_vis_5gev}
\end{figure}

\begin{figure}[htbp]
  \centering
  \includegraphics[width=0.48\linewidth]{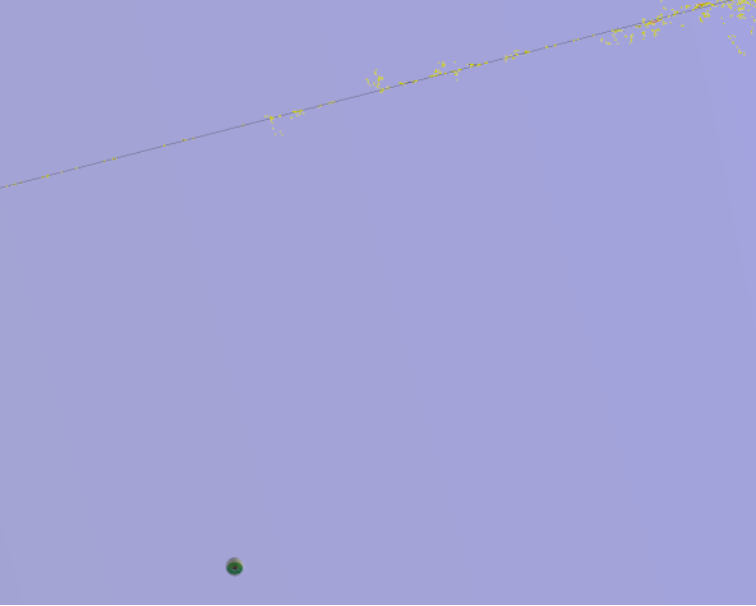}\hfill
  \includegraphics[width=0.48\linewidth]{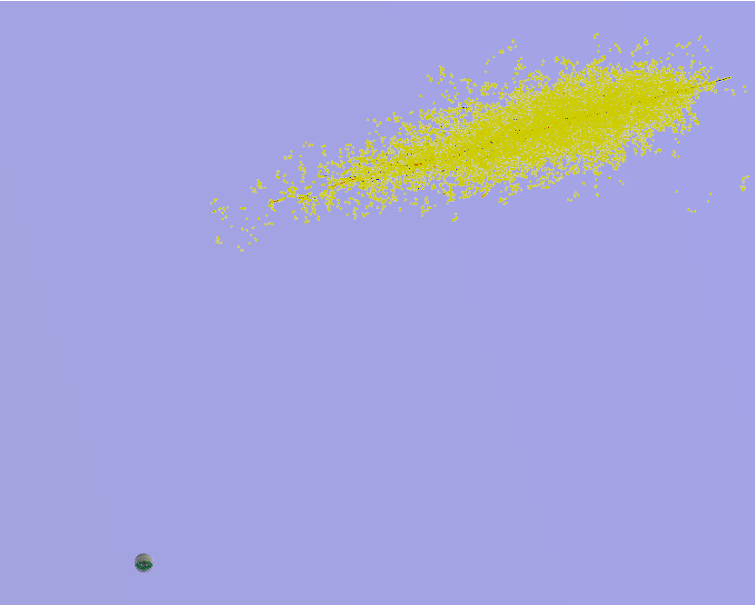}
  \caption{
  \Geant\ \texttt{OMSim} event display of a 100~GeV muon (left) and electron (right) near a DOM.
  Yellow points indicate the sampled trajectory steps of the charged particles (injected lepton and secondaries), Cherenkov photons are not shown.
  }
  \label{fig:geant4_vis_100gev}
\end{figure}
\begin{figure}[htbp]
  \centering
  \includegraphics[width=0.85\linewidth]{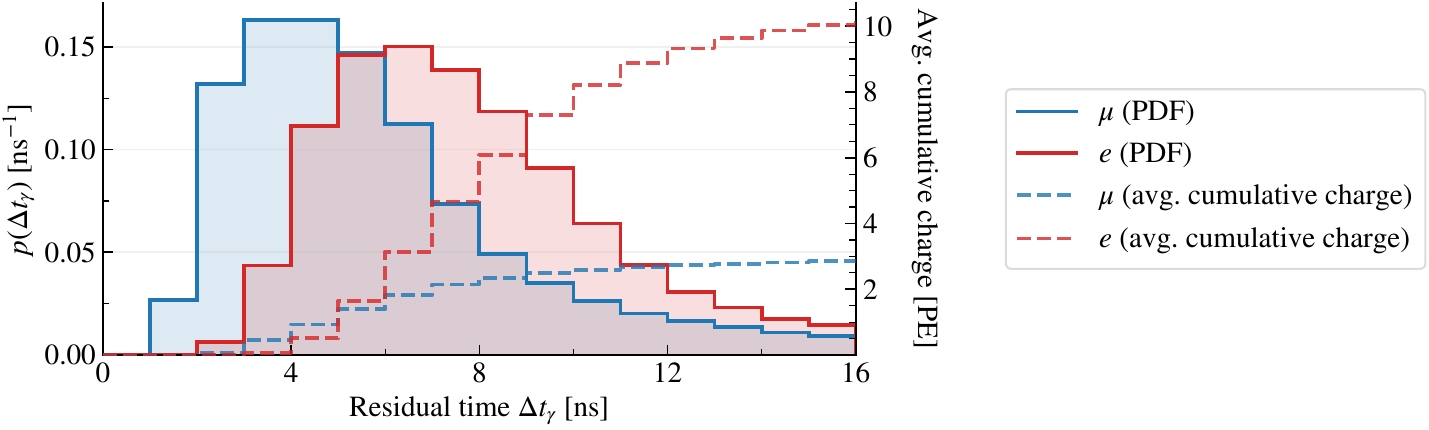}
  \caption{
  Photon density $p(\Delta t_\gamma)$ for muons and electrons at an impact 
  parameter of 20~m and 50~GeV energy from \Geant\ simulations. Dashed curves, read on the right axis, show the corresponding average cumulative detected charge. Both 
  histograms are evaluated in identical $1~\mathrm{ns}$ bins.
  }
  \label{fig:patd_pdf}
\end{figure}

When comparing shapes, we use the normalized distribution
\begin{equation}
p(\Delta t_\gamma) \equiv \frac{1}{N_\gamma}\,\frac{dN_\gamma}{d\Delta t_\gamma} \, ,
\end{equation}
where $N_\gamma$ is the total number of detected photons for a given energy 
and distance configuration and $\Delta t_\gamma$ is the residual time. This 
normalization is used to compare the shape of the photon arrival time 
distributions in \cref{fig:patd_parent} independently of the total photon 
yield, which differs between muon and electron events. Fig.~\ref{fig:patd_pdf} 
shows the normalized probability density for the first 16~ns at a fixed 
impact parameter of 20~m. The muon distribution is more sharply peaked at 
early times, while the electron distribution is broader, reflecting the 
extended shower development.

\begin{figure}[htbp]
  \centering
  \includegraphics[width=0.99\linewidth]{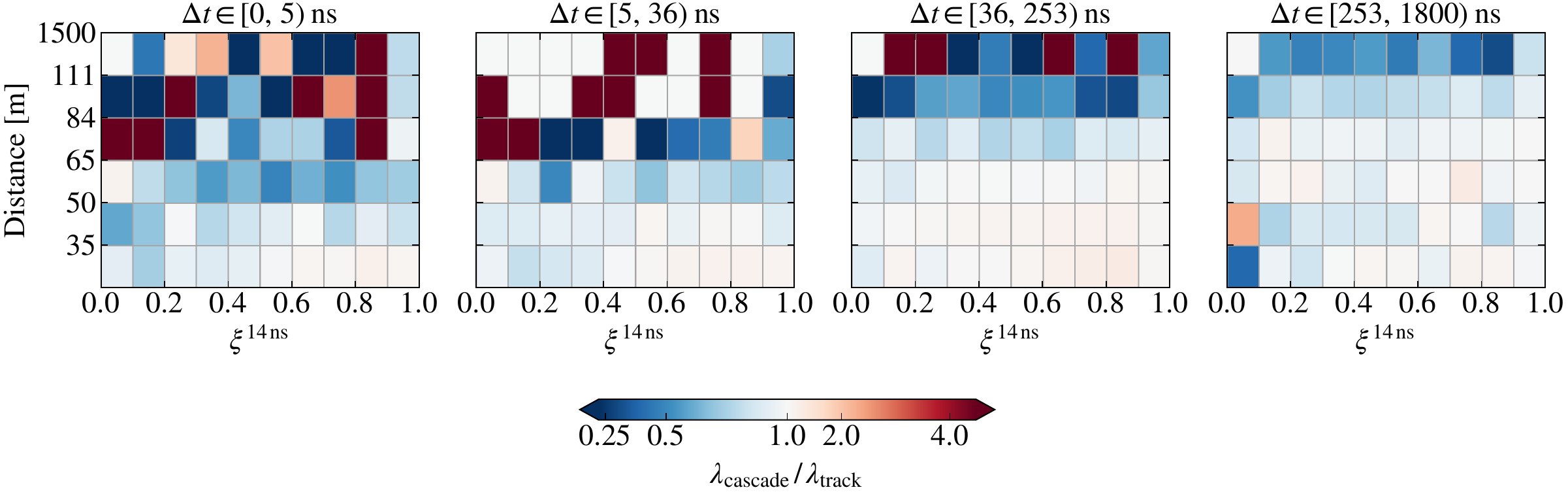}
  \caption{Template ratio 
  $\lambda_\mathrm{cascade}/\lambda_\mathrm{track}$ in each 
  $(\xi^{14\,\mathrm{ns}},\, r,\, \Delta t)$ bin. Blue (red) indicates 
  track-like (cascade-like) regions; white corresponds to a ratio of unity 
  or to bins regularized by the floor $\varepsilon_\lambda$. Discrimination 
  is concentrated in the early-time bin ($\Delta t < 5$~ns) at high 
  $\xi^{14\,\mathrm{ns}}$ and intermediate distances.
  }
  \label{fig:template_ratio}
\end{figure}

\paragraph{Template structure.}
Fig.~\ref{fig:template_ratio} shows the resulting template ratio across the full 
$(\xi^{14\,\mathrm{ns}},\, r,\, \Delta t)$ feature space, illustrating which 
regions of the per-DOM observable space drive the track--cascade separation.

\begin{figure}[htbp]
  \centering
  \includegraphics[width=0.65\linewidth]{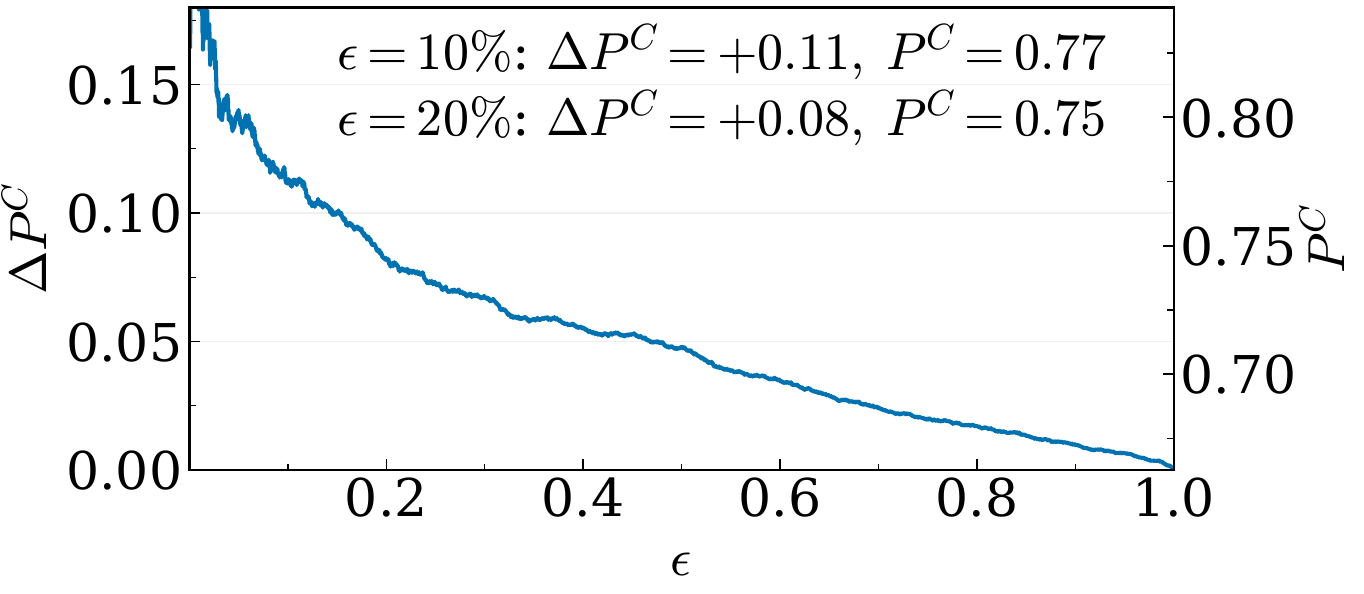}
  \caption{Cascade purity $P^{\mathrm{C}}$ and 
  purity gain $\Delta P^{\mathrm{C}}$ as functions of selection efficiency 
  $\epsilon$ within the \DynEdge\ cascade PID bin, for events with 
  $n_{\mathrm{hit}} \in [10,20)$.}
  \label{fig:purity_boost}
\end{figure}

\paragraph{Purity increase.}
Fig.~\ref{fig:purity_boost} shows the cascade purity $P^{\mathrm{C}}$ and purity 
gain $\Delta P^{\mathrm{C}}$ on the nominal MC set as functions of selection efficiency $\epsilon$, 
within the \DynEdge\ cascade PID bin for events with $n_{\mathrm{hit}} \in 
[10, 20)$. The purity gain is largest at low efficiency and decreases as the 
threshold is relaxed, reflecting the concentration of discriminating power at 
high \WavePID\ TS values.

\paragraph{Computational requirements.}
\WavePID\ is computationally inexpensive. Template construction is performed once from MC and requires storing a
three-dimensional histogram with $6 \times 10 \times 4 = 240$ bins per class,
corresponding to the 6 distance bins, 10 early-charge-fraction bins, and
4 inter-module time bins of \cref{eq:binning}. At inference time, computing 
the test statistic for a single event involves mapping each DOM's features 
to a bin index and summing the corresponding log-odds. This is an 
$\mathcal{O}(n_{\mathrm{hit}})$ operation requiring no matrix multiplications 
or GPU acceleration. On a single CPU core, \WavePID\ processes 
$\mathcal{O}(10^5)$ events per second, making it suitable for both offline 
analysis and potential online filtering applications. The memory footprint 
of the stored templates is less than 10~kB.

\paragraph{Hyperparameter sensitivity.}
The time window $t$ and bin edges $\mathbf{e}_i$ are set empirically, and 
the optimal configuration depends on the target metric. Maximizing overall 
AUC does not necessarily maximize purity at a specific efficiency, and vice 
versa. The configuration presented here was optimized for AUC in the 
\DynEdge\ cascade bin with $10$--$20$ hits, yielding a purity increase from
$66\%$ to $75\%$ at $20\%$ selection efficiency. Alternative configurations that 
prioritize purity over broad classification performance can yield higher purity at fixed efficiency, at the cost of reduced AUC.
\end{document}